\documentclass[useAMS,usenatbib]{mn2e}
\usepackage{graphicx}
\usepackage{lscape}
\usepackage{psfig}

\title{\textit{Swift} Observations of X-ray supernovae}
\author[K.L. Li, Chun.S.J. Pun]{K.L. Li$^{1}$, Chun.S.J. Pun$^{1}$\\
$^{1}$The University of Hong Kong}
\begin{document}

\date{\today}

\pagerange{\pageref{firstpage}--\pageref{lastpage}} \pubyear{2010}

\maketitle

\label{firstpage}

\begin{abstract}
We present a result of X-ray supernovae (SNe) survey using the \textit{Swift} satellite public archive. An automatic searching program was designed to search X-ray SNe among all of the \textit{Swift} archival observations between November 2004 and February 2011. Using the C++ program, 24 X-ray detectable supernovae have been found in the archive and 3 of them were newly-discovered in X-rays which are SN 1986L, SN 2003lx, and SN 2007od. In addition, SN 2003lx is a Type Ia supernova which may be the second X-ray detectable Type Ia after SN 2005ke \citep{2005ke}. Calibrated data of luminous type Ib/c supernovae was consistent to the X-ray emission model done by \citet {inter2}. Statistics about the luminosities and hardness ratio have been done to purpose of getting the X-ray emission features of the X-ray supernovae. The results from this work help investigating the X-ray evolution of SNe and developing similar X-ray SNe surveys in various X-rays missions. 
\end{abstract}


\section{Introduction}
\subsection{X-ray Supernovae}
Supernovae (SNe) are stellar explosions which are luminous in various frequencies of electromagnetic wave. Nowadays, the detection rate of SNe in Optical is about few hundreds per year, while number of SNe detection in X-rays are still rare which is less than 50 detections in total. There are two kinds of X-ray emission mechanisms in SNe which are circumstellar-shock interaction or shock breakout. In shock breakout case, shock breakouts envelope of the progenitor which emits strong X-rays in soft band for a short period of time, says minutes. X-ray transient XRO 080109 or SN 2008D located in NGC 2770 (see \citet{2008D}) is the only case discovered. In circumstellar interaction, the X-ray emission is relatively stable which can be last for few years.  Most of the radiations come from the explosion shock wave pair (i.e. "forward shock" and "reverse shock"). When shock from the SN center interacts with circumstellar matter (CSM) deposited by stellar wind from the progenitors/progenitors' companion star, X-ray photons are produced due to thermal, inverse Compton and synchrotron emissions \citep{2005ke, CF06}. According to the circumstellar interaction model by \citet{inter2}, cooling shell and forward/reverse shocks pair are formed as the ejecta collides with the CSM. In most cases, the forward shocks heat up the CSM to emit hard X-ray (${\ga}$ 10 keV) while the reverse shocks interact with the ejecta to emit soft X-ray (${\la}$ 5 keV). In the early time, the X-ray opaque cool dense shell between the shocks blocks the soft X-ray from the reverse shocked region. After several days or weeks, the expanding cool shell becomes optically thin which allows the pass through of the soft X-ray. Since there is a relatively high density in the reverse shocked region, soft X-ray is strong and therefore dominates the entire X-ray emission. One of the oldest X-ray supernova detected was SN 1970G by a \textit{Chandra} deep image of M101 in 2005, 35 years after the explosion \citep{Universe_in_X-rays}. The luminosity of this SN was $\sim 10^{37}\mathrm{erg\,s^{-1}}$ in 0.3--2 keV which is consistent with a mass loss rate of $2.6\:(v_{w}/\mathrm{10\,km\,s^{-1}}) (M_{\sun}/\,\mathrm{yr\,s^{-1}})$ \citep{1970G}. Recent study indicated that there were even older X-ray SNe: SN 1941C, SN 1959D and SN 1968D \citep{old_sne} with X-ray luminosity around $10^{37-38}\mathrm{erg\,s^{-1}}$ in energy range 0.3--8 keV which were detected by \textit{Chandra}. These old supernovae are important because they fill up the black box between old and young phases of a supernova remnant. 

In this work, X-ray telescope (XRT) on \textit{Swift} was used for searching X-ray SN candidate from different evolution phases. As the \textit{Swift} archival data is plentiful (over 30 observations in different field everyday), it is a suitable place to do such X-ray SNe survey. Total 24 X-ray SNe have been detected in this survey. 3 of them are newly-discovered, i.e. SN 1986L, SN 2003lx and SN 2007od. Scientific statistical results of this survey are presented in \S 2.3 and individual studies of some candidates are discussed in \S 3. 

\section{Results and Data Analysis}

\subsection{Searching Method}
There are over 5000 SNe have been optically discovered since 1885 which were recorded in detail in the Center for Astrophysics (CfA) archive \footnote{http://www.cfa.harvard.edu/iau/lists/Supernovae.html}. Since the sample size of survey target is so large, an automated program was developed to screen all Swift archival data in photon counting mode in the period November 2004 to February 2011 and detect if there is any X-ray SNe in the field. Here gives a brief idea about how the program was working. 

Firstly, we developed a local \textit{Swift} database by downloading all old archival data from the \textit{Swift} public archive and then using a time-based job scheduler \textit{cron} to update fresh calibrated X-ray observations to the database at every midnight. Since the file size of a full data set (i.e. auxiliary files, UVOT data and etc) is large which is usually larger than 10~Mb per observation in size, so only X-ray event files which contain information of arrival photons were downloaded to minimize time and computer resources required. All observations in the database were classified into groups according to their aim points. 

Afterward, A list which contains all detected SNe positions was extracted and updated from CfA. The program then did positional matching between the the list and the database. Since the real field of view of \textit{Swift} is $23\arcmin.6 \times 23\arcmin.6$, we set the critical separation to be half length of the diagonal $r = 16\arcmin.7$. Once the angular separation between a observation and a SN is less than the critical value, the existence of the corresponding SN in the field of view is confirmed.

For observations that contain SN/SNe in their fields of view, full set of data were downloaded. Observations that belong to the same SN were merged by \textit{xselect} to optimize the signal to noise ratio and a corresponding X-ray image in energy range 0.3 to 10 keV was extracted. All extracted images were then processed by \textit{wavdetect} of CIAO to detect all X-ray sources in the field. We set the wavelet scales to be a series from 2 to 8 increasing by a factor of $\sqrt{2}$ , (i.e. 2.0, 2.828, 4.0, 5.657, 8.0). In order to detect those X-ray sources with low counts, the values of the signal and background thresholds were set to ${5\times10^{-6}}$. In addition, exposure maps generated by \textit{xrtexpomap} were used to correct the variation of effective area throughout the image. 

According to the pointing accuracy and the angular resolution of \textit{Swift} XRT, a box search of size 10${\arcsec}$ was used to find any positional coincidence bewteen the corresponding SN and sources in the \textit{wavdetect} source list. Sources with such coincidences became X-ray SN candidate and went through another 10${\arcsec}$ box search in SIMBAD to do source identification. In case a candidate consists of 1 or more non-supernova identifications, it may be considered to be contaminated by the nearby sources and would be eliminated from the candidate list. 

Finally, soft band (0.3--1.1 keV), median band (1.1--2.0 keV) and hard band (2--10 keV) images were extracted from the event files and \textit{wavdetect} with the above parameters were applied to each band images to check the count rates of each band and hardness ratios of the X-ray SN candidates.

\subsection{Justification of the searching method}
The ASDC Swift-XRT Simulator\footnote{http://www.asdc.asi.it/simulator/swift/} was used to simulate \textit{Swift} observations of series of fake X-ray sources. The simulated observations were then processed by the searching routines discussed to give blind searches of the X-ray sources. Percentage of sources recovered should be a key parameter for justifying effectiveness of the program. In this test, an index 2 power law  with hydrogen column density ${\mathrm{n_H}=2.17\times10^{20}\mathrm{cm}^{-2}}$ model was used to generate 64 X-ray sources in different locations in image. We simulated their \textit{Swift} observations with exposure times 2000~seconds and 5000~seconds. In 2000~seconds simulation, 78\% of the X-rays sources were detected by the program while all X-ray sources were correctly identified and only got 3 wrong identifications for the 5000~seconds case. 

\subsection{Searching Result}
24 X-ray SNe candidates were found in the \textit{Swift} public archive using our program in the period between November 2004 and February 2011, including 14 Type II, 9 Type Ib/c and 1 Type Ia X-ray SN candidates (see Table~1). Many candidates have been previously studied including SN 1978K in NGC 1313 \citep{1978K}, 1993J in M81 \citep{1993J_lc} and etc. We compared our result with the previous publications of \textit{Swift} and found that SN 2005ke, 2008M, 2008ax, 2008ij, 2009gj, 2010F, 2010jl, 2010jr \citep{2005ke, 2008ax, 2008ij, 2009gj, 2008M, 2010F, 2010jl, 2010jr} were missing in our list. The reason was mainly believed due to the different approaches of the SN/source detections. The general X-ray SNe searching methods are targeted searches while a blind search method was used in this work. The latter could be done in an automatic way easily. In order to due with a 5 years database (about fifty thousand observations) automatically, blind search approach was used in this project and it was supposed to be the major reason why there is a slightly difference between the results. Indeed, two of the above missing SNe, SN 2010jl and 2010jr, were found in the out blind search. However, they are too close to the host galaxies, so SN 2010jl was heavily contaminated by the galaxy and SN 2010jr is very hard in spectrum which is likely to be something else (i.e. emission from its host galaxy). 

\begin{figure*}
\includegraphics[width=177mm]{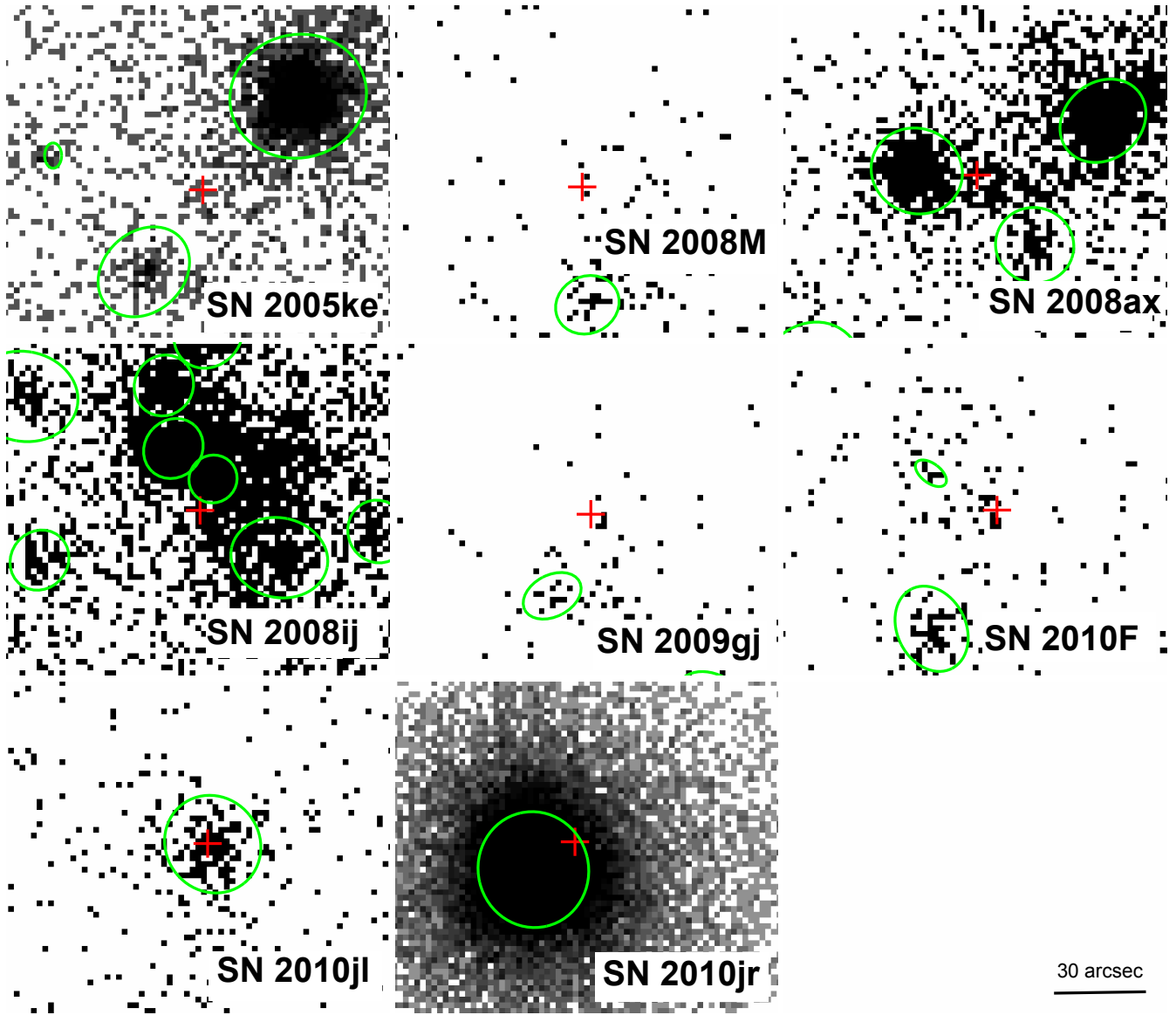}
\caption{The XRT images of energy band 0.3--10 keV of the "missing" \textit{Swift} X-ray supernovae. The green ellipses show the sources detected by \textit{wavdetect} in this work while the red crosses show the corresponding SN locations. 
}
\label{official}
\end{figure*}

\clearpage
\begin{landscape}
\begin{table}
\centering
\begin{minipage}{200mm}
\caption{Complete list of supernova found in \textit{Swift}}

\begin{tabular}{@{}cccccccccccc}
\hline
SN Name & SN RA & SN DEC & OBS DATE & OBS END & EXP & SRC RA & SRC DEC & SEP & SIGN \\ 
 & & & (UT) & (UT) & (s) & & & (arcsec) & (sigma) \\ 
\hline
SN 1970G & 14 03 00.83 & +54 14 32.8 & 2005-01-20 01:23:34 & 2007-04-19 10:26:56 & 65481 & 14 03 00.43 & +54 14 32.1 & 4.1 ${\pm}$ 1.8 & 3.50\\ 
SN 1978K & 03 17 38.62 & -66 33 03.4 & 2006-02-03 22:56:19 & 2009-12-20 02:42:58 & 49940 & 03 17 38.66 & -66 33 02.0 & 1.4 ${\pm}$ 0.6 & 70.92\\ 
SN 1979C & 12 22 58.58 & +15 47 52.7 & 2005-11-06 16:00:55 & 2006-03-29 15:15:58 & 57636 & 12 22 58.59 & +15 47 47.9 & 4.8 ${\pm}$ 0.9 & 8.60\\ 
SN 1986L & 04 17 29.40 & -62 47 04.0 & 2005-08-10 02:45:26 & 2007-12-12 16:42:57 & 50474 & 04 17 29.19 & -62 46 58.9 & 6.0 ${\pm}$ 1.3 & 14.83\\ 
SN 1987A & 05 35 28.01 & -69 16 11.6 & 2005-01-24 01:09:05 & 2006-01-11 09:48:56 & 35728 & 05 35 28.07 & -69 16 10.8 & 1.1 ${\pm}$ 0.3 & 134.90\\ 
SN 1993J & 09 55 24.78 & +69 01 13.7 & 2005-04-21 00:47:46 & 2011-02-17 14:35:58 & 293723 & 09 55 24.64 & +69 01 13.6 & 2.0 ${\pm}$ 0.6 & 31.53\\ 
SN 1996cr & 14 13 10.05 & -65 20 44.8 & 2007-03-23 06:08:28 & 2009-11-16 22:54:55 & 18492 & 14 13 10.04 & -65 20 44.4 & 0.4 ${\pm}$ 0.5 & 35.33\\ 
SN 2003lx & 16 19 21.65 & +41 05 23.8 & 2008-01-10 20:39:16 & 2008-01-14 22:52:57 & 13810 & 16 19 21.70 & +41 05 25.7 & 2.1 ${\pm}$ 1.2 & 17.20\\ 
SN 2005ip & 09 32 06.42 & +08 26 44.4 & 2007-02-14 01:57:14 & 2009-11-19 05:28:58 & 25304 & 09 32 06.44 & +08 26 44.8 & 0.5 ${\pm}$ 0.5 & 32.94\\ 
SN 2005kd & 04 03 16.88 & +71 43 18.9 & 2007-01-24 00:56:52 & 2008-08-21 22:44:58 & 18168 & 04 03 17.19 & +71 43 19.2 & 4.0 ${\pm}$ 2.0 & 12.90\\ 
SN 2005nc & 18 32 32.58 & +26 20 22.6 & 2005-05-25 00:03:00 & 2005-06-28 23:20:57 & 268557 & 18 32 32.62 & +26 20 20.6 & 2.1 ${\pm}$ 0.2 & 51.77\\ 
SN 2006aj & 03 21 39.66 & +16 52 02.1 & 2006-02-18 05:13:38 & 2010-12-15 23:13:57 & 357796 & 03 21 39.69 & +16 52 00.7 & 1.4 ${\pm}$ 0.2 & 87.06\\ 
SN 2006bp & 11 53 55.74 & +52 21 09.4 & 2006-04-10 12:51:21 & 2007-03-31 22:10:56 & 87892 & 11 53 55.84 & +52 21 13.3 & 4.0 ${\pm}$ 1.2 & 3.51\\ 
SN 2006jc & 09 17 20.78 & +41 54 32.7 & 2006-10-13 16:10:14 & 2011-01-04 22:02:56 & 202441 & 09 17 20.90 & +41 54 37.6 & 4.9 ${\pm}$ 0.4 & 34.98\\ 
SN 2006jd & 08 02 07.43 & +00 48 31.5 & 2007-11-16 17:35:21 & 2010-06-09 03:07:56 & 69347 & 08 02 07.36 & +00 48 31.7 & 0.7 ${\pm}$ 0.4 & 39.68\\ 
SN 2007od & 23 55 48.68 & +18 24 54.8 & 2007-11-05 16:13:08 & 2008-08-23 19:02:56 & 65929 & 23 55 48.69 & +18 24 52.1 & 2.7 ${\pm}$ 1.3 & 4.21\\ 
SN 2007uy & 09 09 35.40 & +33 07 09.9 & 2008-01-06 00:33:07 & 2009-02-21 23:01:55 & 334518 & 09 09 35.02 & +33 07 09.0 & 1.1 ${\pm}$ 0.7 & 5.82\\ 
SN 2008bo & 18 19 54.41 & +74 34 21.0 & 2008-04-04 14:54:02 & 2009-02-14 23:05:57 & 383873 & 18 19 54.60 & +74 34 20.7 & 2.0 ${\pm}$ 0.6 & 34.77\\ 
SN 2008D & 09 09 30.62 & +33 08 20.1 & 2008-01-06 00:33:07 & 2009-02-21 23:01:55 & 334518 & 09 09 30.62 & +33 08 20.2 & 0.0 ${\pm}$ 0.3 & 46.80\\ 
SN 2008hw & 22 39 50.39 & -40 08 49.1 & 2008-10-07 05:26:52 & 2008-10-27 14:00:56 & 188541 & 22 39 50.59 & -40 08 47.0 & 3.1 ${\pm}$ 0.3 & 85.51\\ 
SN 2009dd & 12 05 34.10 & +50 32 18.6 & 2009-04-15 14:50:05 & 2009-09-21 23:59:57 & 104014 & 12 05 34.58 & +50 32 20.7 & 7.3 ${\pm}$ 1.0 & 8.85\\ 
SN 2009mk & 00 06 21.37 & -41 28 59.8 & 2009-12-17 03:30:58 & 2010-01-30 23:59:57 & 38301 & 00 06 20.72  & -41 29 03.8  & 8.8 ${\pm}$ 1.4 & 1.90\\ 
SN 2009nz & 02 26 19.90 & -18 57 09.0  & 2009-11-28 01:02:16 & 2010-06-10 23:44:57 & 470379 & 02 26 20.03  & -18 57 07.2  & 2.6 ${\pm}$ 0.1 & 154.47\\ 
SN 2010bh & 07 10 31.80  & -56 15 20.2  & 2010-03-16 21:53:21 & 2010-03-23 20:55:33 & 52488 & 07 10 30.95  & -56 15 17.5  & 12.5 ${\pm}$ 1.2 & 13.18\\ 
\hline
\end{tabular}
\end{minipage}
\end{table}
\end{landscape}

\subsection{Data Reduction}
We used HEAsoft of version 6.6.2 to do the data analysis. Tasks \textit{xrtgrblc} and \textit{xrtgrblcspec} were used to extract lightcurves and spectra while the \textit{Swift}-XRT data products generator \citep{autometh} was used\footnote{http://swift.ac.uk/user\_objects} to do some crosschecks sometimes. We binned all spectral products with 20 counts per bin at first. Those binned spectra with 4 or less data points were re-binned with 10 counts per bin. Re-binned spectra with 7 or less data points were re-binned with 5 counts per bin. Spectra with 9 or less data points after 5 counts binning were regarded as low signal to noise which were not used to do model fitting. The background regions were chosen automatically by the tasks. Source-free background regions would be extracted by human sometimes to avoid contamination from some nearby bright X-ray sources or inappropriate background selections at the edges of images. Finally, XSPEC was used to do the spectral fitting to examine physical parameters of the sources. Basically, combinations of simple photon power law model and Raymond-Smith thermal plasma model \citep{raymond} were primary models of the analysis. Column densities from Leiden/Argentine/Bonn (LAB) Survey of Galactic HI were used as the Galactic $\mathrm{n_H}$, and $\mathrm{n_H}$ excess components would be added if necessary. The best fit results of chi-square fittings of 16 X-ray SN candidates are shown in the Table \ref{spectra_SNe}. 

\subsection{Properties of X-ray supernovae spectra}
\begin{figure}
\includegraphics[width=84mm]{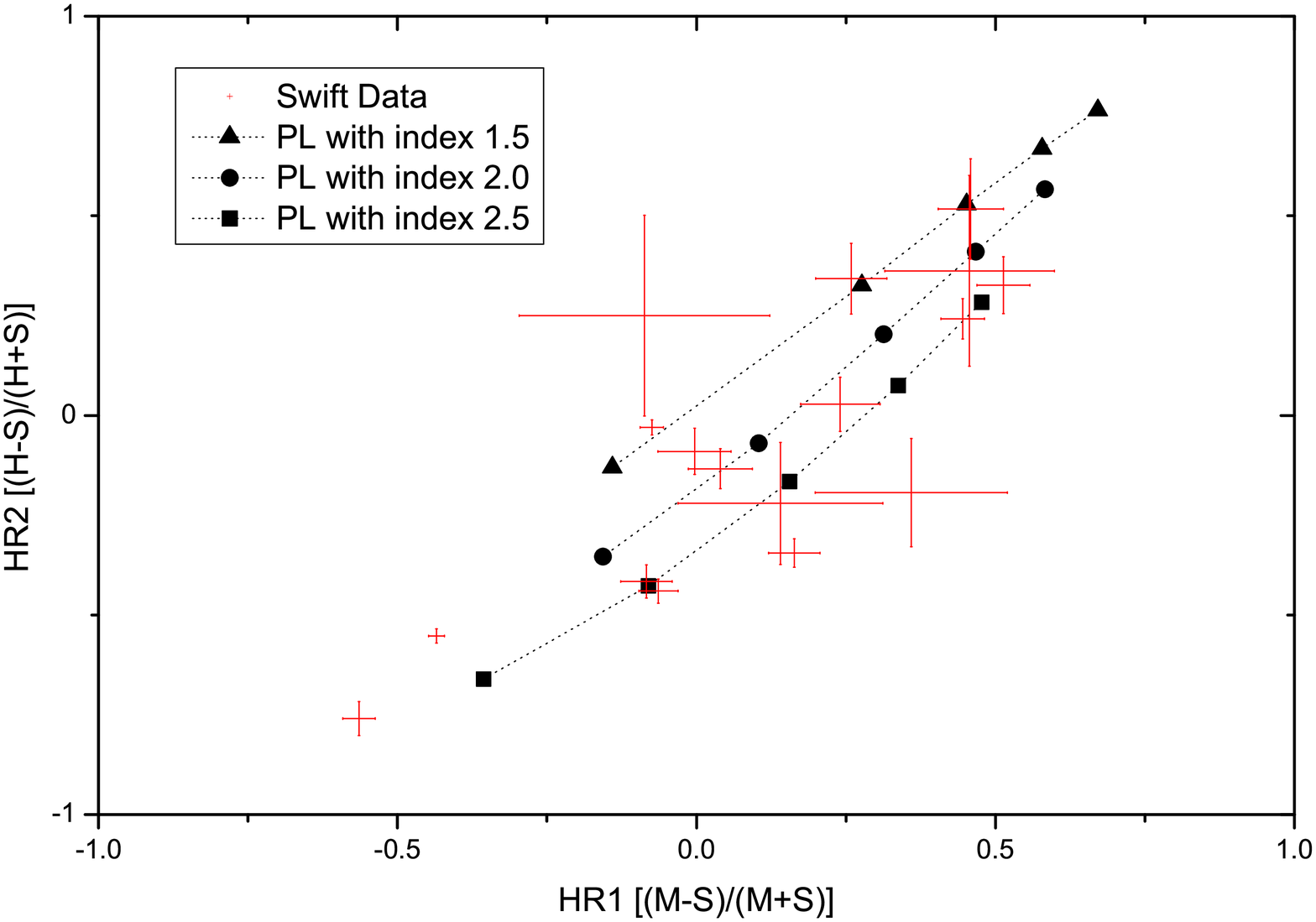}
\caption{Hardness diagram of the X-ray SNe. The points are the PIMMS of version 4.1a predicted values using Power Law (PL) with different values of column density (from left: 0.1, 1, 2, 3, 4$\times 10^{21} \mathrm{cm}^{-2}$).  The soft band (S: 0.3--1.0 keV), median band (M: 1.0--2.0 keV), and hard band (H: 2.0--10 keV) hardness ratios $\mathrm{HR_1}$ and $\mathrm{HR_2}$ are plotted.}
\label{Hardness Ratio}
\end{figure}

In the early phase of the explosion, the strong photosphere radiation is Compton scattered which forms a high energy tail to the photosphere emission. Each photon increases its energy by a factor $\triangle\nu/\nu = 4kT_\mathrm{e}/m_\mathrm{e}c^2 \ga 1$ where $m_e$ and $T_e$ are electron mass and electron gas temperature, which create a power law spectrum with a photon index \citep{fransson_82,Supernovae and Gamma-Ray Bursters}
\begin{equation}
\alpha = \lbrace \frac{9}{4}- \frac{m_\mathrm{e}c^2}{kT_\mathrm{e}} \ln [\frac{\tau_\mathrm{e}}{2}(0.9228- \ln \tau_\mathrm{e})] \rbrace ^{\frac{1}{2}} - \frac{3}{2}, 
\end{equation}
where $\tau_e$ is the optical depth to electron scattering behind the circumstellar shock. The typical value is $1 \la \alpha \la 3$ \citep{fransson_82}. In order to compare the SNe data to the model prediction, hardness ratio diagram of high signal (count $>$ 50) X-ray supernovae candidates was plotted (Table \ref{Hardness Ratio}). In the diagram, hardness ratios $\mathrm{HR_1=\frac{M-S}{M+S}}$ and $\mathrm{HR_2=\frac{H-S}{H+S}}$  were plotted where S = soft band (S: 0.3--1.0 keV), M = median band (M: 1.0--2.0 keV), and H = hard band (H: 2.0--10 keV). With the reference trends from the Portable Interactive Multi-Mission Simulator (PIMMS) prediction of power law model with various photon indexes and column densities, the diagram shows clearly that the ratio between the hardness HR1 and HR2 are consistent with a power law model of index $\sim2$ which is consistent the prediction. Spectral fitting resuls also agreed with the theoretical model. As shown in the Table \ref{spectra_SNe}, most of the supernovae (except SN 1993J) show non-thermal components (power law) in their spectra with a mean photon index $\sim2$ while some spectra also show thermal components with plasma temperatures about $\sim0.8$ keV. If ages of SNe were taken into the account, it is clear that older supernovae spectra were fit well with thermal plus non-thermal model while younger supernovae could be satisfied with single power law. One explanation is that non-thermal blast shock emissions dominated the whole X-ray emissions of supernovae which created single power law continuum in the early phase. After years, the shock velocities decreased as the swept-up mass behind the circumstellar shock increased which weakened the non-thermal part, hence the thermal emissions (and the line emissions from elements) started to be significant. 

\begin{table*}
\centering
\caption{Spectral properties of X-ray SNe candidates}
\begin{tabular}{@{}lcccccc}
\hline
SN Name & Distance & $\mathrm{n_H}$ (Galactic) & $\mathrm{n_H}$ (Intrinsic) & Photon Index & Plasma Temperature (kT) & Reduced-$\chi^2$\\ 
& (Mpc) & ($10^{20}\,\mathrm{cm}^{-2}$) & ($10^{20}\,\mathrm{cm}^{-2}$) & & ($\mathrm{keV}$) & \\ 
\hline
SN 1978K & 4.5 & 4.07 & $24^{+9}_{-8}$ & $2.7^{+0.4}_{-0.4}$ & $0.9^{+0.1}_{-0.1}$ & 1.08 \\ 
SN 1986L & 16 & 2.17 & - & $1.4^{+0.2}_{-0.2}$ & $0.7^{+0.2}_{-0.2}$ & 1.25 \\ 
SN 1987A & 0.05 & 26.5 & - & $3.8^{+0.1}_{-0.1}$ & $0.7^{+0.1}_{-0.1}$ & 1.42 \\ 
SN 1993J & 3.6 & 5.57 & - & - & $0.8^{+0.1}_{-0.1}$ & 1.04 \\ 
SN 1996cr & 17 & 55.6 & - & $1.1^{+0.2}_{-0.2}$ & - & 1.27 \\ 
SN 2005ip & 30 & 3.56 & - &  $0.5^{+0.2}_{-0.2}$ & - & 1.16 \\ 
SN 2005nc/GRB050525A$^b$ & 1780 & 9.07 & - & $1.7^{+0.1}_{-0.1}$ & - & 1.04 \\ 
SN 2006aj/GRB060218$^b$ & 136 & 9.37 & $46^{+6}_{-10}$ & $3.8^{+0.4}_{-0.3}$ & $0.7^{+0.1}_{-0.2}$ & 1.58 \\ 
SN 2006jc & 23.6 & 1.51 & $19^{+13}_{-7}$ & $1.7^{+0.2}_{-0.2}$ & - & 0.81 \\ 
SN 2006jd & 79 & 4.52 &  $9.1^{+11}_{-7}$ & $1.2^{+0.2}_{-0.2}$ & - & 1.33 \\ 
SN 2008bo & 21 & 5.37 & - & $1.9^{+0.2}_{-0.2}$ & - & 1.16 \\ 
SN 2008D$^a$ & 27 & 1.7 & - & $1.5^{+0.6}_{-0.5}$ & - & 0.37 \\ 
SN 2008hw/GRB081007$^b$ & 1620 & 1.38 & $56^{+6}_{-10}$ & $2.1^{+0.2}_{-0.1}$ & - & 0.84 \\ 
SN 2009dd & 13 & 1.83 & - & $1.9^{+0.6}_{-0.6}$ & - & 1.93 \\ 
SN 2009nz/GRB091127$^b$ & 239 & 2.8 & 26 & $2.4^{+0.1}_{-0.1}$ & - & 1.2 \\ 
SN 2010bh/GRB100316D$^b$ & 1530 & 7 & - & $1.7^{+0.6}_{-0.6}$ & - & 0.82 \\ 
\hline
\textit{a. Spectrum after outburst was used.}\\
\textit{b. Time average spectra were used.}
\end{tabular}
\label{spectra_SNe}
\end{table*}


\subsection{X-ray Supernovae luminosity}
\begin{figure}
\includegraphics[width=84mm]{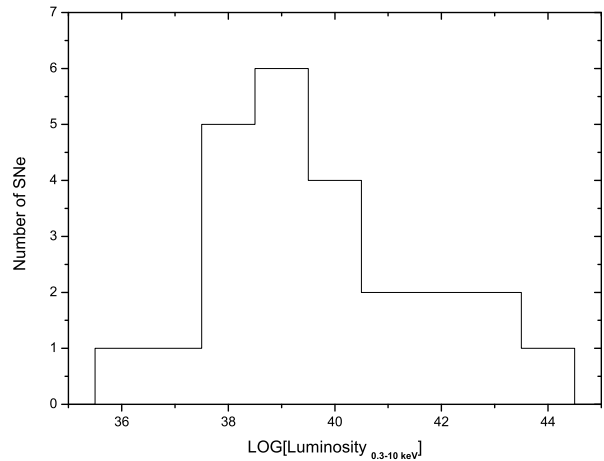}
\caption{The observed X-ray 0.3--10 keV luminosities of supernovae calculated by PIMMS, from SN 1987A (blue shadow) at $\mathrm{log}_{10}(\mathrm{L}_{0.3-10 \mathrm{keV}}) = 36.3$ to SN-GRB candidate SN 2008hw at $\mathrm{log}_{10}(\mathrm{L}_{0.3-10 \mathrm{keV}}) = 44.2$, with an average peak luminosity $\mathrm{log}_{10}(\mathrm{L}_{0.3-10 \mathrm{keV}}) = 40.3$, reflecting partly the different stages of evolution of the observed supernovae.}
\label{luminosity function}
\end{figure}

We estimated the X-ray luminosity of the X-ray supernova candidates by the spectral model fitting. For those low signal-to-noise candidates data, the net count rates with power law of photon index 2 were used to estimate the flux using PIMMS. Fig. \ref{luminosity function} shows the luminosity function of all candidates. Most of candidates have a luminosity of $\sim 10^{40}\mathrm{erg/s}$ in energy band 0.3--10 keV. The most luminous SNe were the GRB-SN associations which have a luminosity of $\sim 10^{44}\mathrm{erg/s}$ while the dimmest one was $\sim 10^{36}\mathrm{erg/s}$, which belongs to the SNR 1987A in Large Magellanic Cloud. This wide X-ray luminosity range partly reflects the different stages of evolution of the observed supernovae. 

In addition to know more the properties of the SNe progenitors, we used the X-ray luminosities to estimate the CSM density profile. \citet{inter2} suggested free-free luminosity from reverse shock can be estimated from
\begin{equation}
L_{\mathrm{rev}} = \frac{\pi{R}^2\,n_\mathrm{e}(R)\,n_\mathrm{i}(R)\,\Lambda(T)\,t\,V}{1 + [2n_\mathrm{e}(R)\,\Lambda(T)/(m_\mathrm{p}\,{V}^2)]\,t}\:\mathrm{erg\,s^{-1}},
\end{equation}
where $R$ and $T$ are radius and temperature of the reverse shock, $n_e$ and $n_i$ are the number densities of electrons and ions, $t$ is age of the supernova, $V$ is the shock velocity, $m_p$ is the proton mass and $\Lambda$ is the cooling factor, which can be expressed as $\Lambda = 6.2 \times 10^{-19}\,T^{-0.6}\:\mathrm{erg\,cm^{3}\,s^{-1}}$ for $10^5\,\mathrm{K} < T \lid 4 \times 10^7\,\mathrm{K}$. With the assumption of the CSM is hydrogen and $R_2 = V_\mathrm{ej}\,t = (n-2)\,V_2\,t$, we get
\begin{equation}
L_{\mathrm{rev}} = \frac{4\pi{R}^3\,{\rho_\mathrm{rev}}^2\,\mu_\mathrm{e}^2\Lambda}{2{\mu_\mathrm{H}}^2{m_{\mathrm{e}}}^2(n-2)\lbrace 1 + \frac{4\Lambda\rho_\mathrm{rev}\mu_\mathrm{e}\,(n-2)^2t}{m_\mathrm{p}{V_\mathrm{ej}}^2\mu_\mathrm{H}m_{\mathrm{e}}} \rbrace}\:\mathrm{erg\,s^{-1}},  
\end{equation}
where $\rho_{cs}$ and $\rho_{rev}$ refers to the densities behind the circumstellar shock and the reverse shock in unit of $g\,cm^{-3}$. We assumed the progenitors are Wolf-Rayet stars with a mass loss rate $\dot{M} = 2 \times 10^{-4} \, \mathrm{M_{\sun} \, yr^{-1}}$ and $w_{\mathrm{wind}} = 1000\,\mathrm{km\,s^{-1}}$. Some typical values $n = 15$ and $T = 10^7\,\mathrm{K}$ (which is about 0.7 keV which is consistent to our spectral fitting result) and $V_{\mathrm{ej}} = 35000\,\mathrm{km\,s^{-1}}$ were also applied in the calculation. With the relation $\rho_{\mathrm{rev}}/\rho_{\mathrm{cs}}\approx 0.602\,(n-3)^2$ and the X-ray luminosities of various Type Ib/c SNe, we get a typical picture of Type Ib/c CSM density profile. After comparing it with a steady flow model, $\rho_\mathrm{cs} = \dot{M}/\pi w_{\mathrm{wind}}r^2$ \citep{inter2}, the difference between them increases as time/radius raises. If a dropping temperature (i.e. $T \propto t^{-1.3}$) was used , the result then matched perfectly with the prediction of the steady flow model. Although the calculation looks perfect, it is important to notice that the decay rate $T \propto t^{-1.3}$ is not significant to the real situation since there are many assumptions and fixed parameters (i.e. the mass loss and the shock velocity) in the model. The decay rate would be different if other parameters were freed. 

\section{Study of individual X-ray Supernova}

\subsection{SN 1986L}
SN 1986L was discovered on October 7.6, 1986, with an apparent magnitude of 13.7 in the location of 50${\arcsec}$ due west of the NGC1559's nucleus (R.A. = 4h17m0, Decl. = -62deg55${\arcmin}$, equinox 1950.0) \citep{1986L1,1986L2}. It is located at NGC1559 where is 16 Mpc from the earth \citep{SN_distance}. SN 1986L was classified as a core-collapse Type II supernova.

\begin{figure*}
\includegraphics[width=177mm]{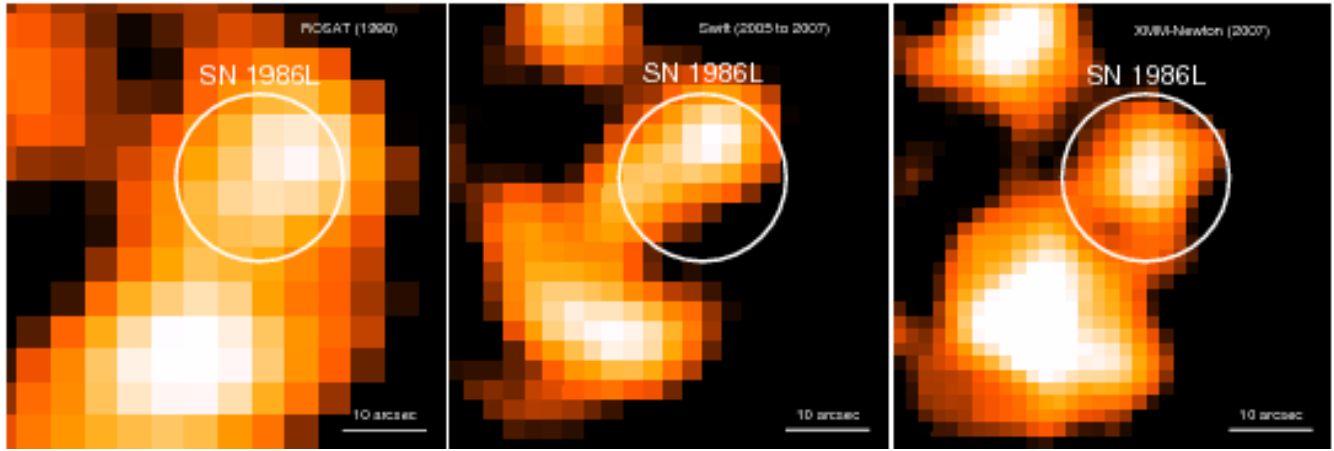}
\caption{Aligned soft (0.3--2.4 keV) X-ray images of SN 1986L taken by \textit{ROSAT} (Left), \textit{Swift} (Middle) and \textit{XMM-Newton} (Right) with the position of the supernova indicated by a 10 arcsec radius green circle. The slightly different positions of the X-ray source observed are consistent with the astrometric accuracies of the different satellites. }
\label{SN 1986L XMM, hri and swift}
\end{figure*}

\textit{Swift} did several observations on the field of SN 1986L from 2005-08-10 to 2007-12-12 and detected an excess of X-ray counts (0.3--10 keV) located on 7.1${\arcsec}$ from the optical SN position at a 15.4\,${\sigma}$ level in the 50.5\,ks combined image. The XRT net count rate is ${(2.5\pm0.3)\times10^{-3}}$ count/s. By assuming the column density ${\mathrm{n_H}=2.17\times10^{20}\mathrm{cm}^{-2}}$ and its distance from the earth was ${\mathrm{Mpc}}$, the luminosity was approximately ${(1.8_{-0.4}^{+0.2})\times10^{40}\mathrm{erg\,s^{-1}}}$ in 0.3--10 keV band.

The \textit{ROSAT} High Resolution Imager (HRI) and \textit{XMM-Newton} observed at SN 1986L on 1990-06-01 and 2007-04-26 with exposure time 17.5ks and 12.5ks, respectively (Fig. \ref{SN 1986L XMM, hri and swift}). In the figure, A position coincidental X-ray source was observed in all \textit{ROSAT}, \textit{Swift} data and \textit{XMM-Newton} data. The sources seem to have different locations in each mission due to the different position accuracies (i.e. ${\sim6\arcsec}$ for \textit{ROSAT}; ${\sim5\arcsec}$ for \textit{Swift}; ${\sim2\arcsec}$ for \textit{XMM-Newton}). The low exposure time of \textit{ROSAT} observation brought a low signal-to-noise ratio image which is not much informative. Using the data of \textit{XMM-Newton}, the luminosity in energy band 0.3--10 keV was estimated which is ${(3.1_{-1.3}^{+0.9})\times10^{39}\mathrm{erg\,s^{-1}}}$ in 2007.


\subsection{SN 1987A}
\begin{figure}
\includegraphics[width=84mm]{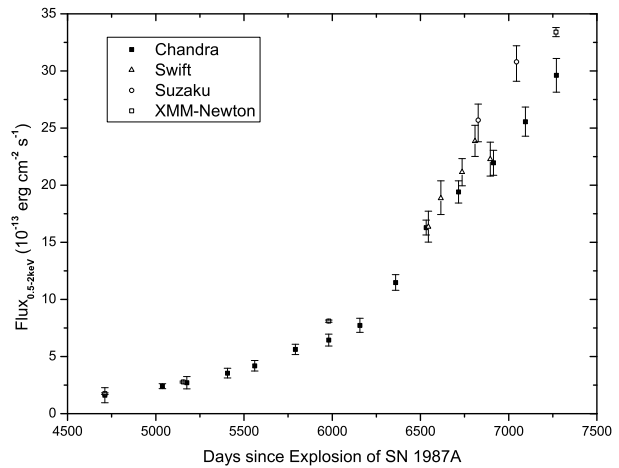}
\caption{X-ray light curve of SN 1987A in the 0.5-2.0 keV band for the past $\sim10$ years. The XMM-Newton flux values are from \citet{1987A_XMM} (for the first three data) and \citet{1987A_XMM2} (for the last data). The Chandra flux values are from \citet{1987A_chan_lc}. The Swift fluxes are from this work, which were estimated by the XSPEC model wabs*(vpshock+vpshock) with temperatures of 0.3 and 2.3 keV, and abundances from \citet{1987A_abund}. 
}
\label{SN1987A X-ray lc Image}
\end{figure}

SN 1987A is one of the most famous and well-known supernovae in astronomical history because it is the only supernova that was detected in neutrino before optical light reached the earth till now. SN 1987A is located in the Large Magellanic Cloud, which is 50 kpc from us. This short distance allows us to resolve the detailed structure of the entire debris in optical. Apart from the optical features, the SN 1987A is distinctive because it is the only Type IIP supernova that was detected in X-rays. It is also the X-ray dimmest supernova ($\sim 10^{36}$ erg s$^{-1}$ in 0.3--10keV) ever detected, which is the result of the short distance between LMC and the earth. \citep{Universe_in_X-rays}. The soft band (0.5-2.0 keV) light curve of the SN 1987A in the past $\sim$10 years (Fig. \ref{SN1987A X-ray lc Image}) shows that an exponential increasing trend almost from $\sim$5500 days after the outburst. This rapid increase was due to the interaction between the blast shock wave and the medium in the inner ring. 

\textit{Swift} observed SN 1987A six times between April 2005 and January 2006. Using a double constant temperature plane-parallel shock plasma model (i.e. wabs*vpshock*vpshock) with hydrogen column density and elemental abundances from the best fit values of \citet{1987A_abund}, the best fit shock temperatures are $0.41^{+0.10}_{-0.10}$~keV and $2.87^{+0.60}_{-0.74}$~keV respectively which are consist with results $0.41^{+0.60}_{-0.25}$~keV and $3.40^{+0.77}_{-0.66}$~keV of \citep{1987A_abund}. The estimated luminosity of 0.5--2.0 keV is $L = 1.40^{+0.04}_{-0.04} \times 10^{36}\mathrm{\,erg\,s{^{-1}}}$ which is also consistent to the trend of the multi-mission lightcurve of SN 1987A which is shown as Fig. \ref{SN1987A X-ray lc Image}. 



\subsection{SN 1993J}
SN 1993J was discovered on March 28, 1993 in the nearby spiral galaxy M81 \citep{1993J}. It is the most intensively monitored X-ray SN after SN 1987A. The first X-ray observation was done by ASCA 6 days after its discovery. Afterward, the evolution of the SN was monitored by ASCA, ROSAT (6-1800 days); \textit{XMM-Newton}, \textit{Chandra} (2500-5500 days); \textit{Swift} (after 4800 days) in the past 16 years. With the substantial number of complete observations, the entire SN evolution analysis could be performed through spectral fitting.

\begin{figure}
\includegraphics[width=84mm]{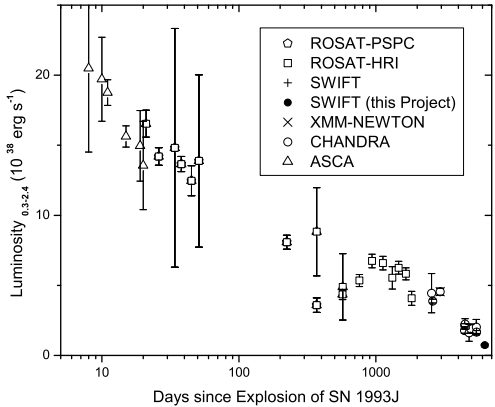}
\caption{0.3--2.4 keV X-ray lightcurve compiled from ASCA: triangles; ROSAT: squares and pentangles; \textit{Chandra}: circles; \textit{XMM-Newton}: crosses; \textit{Swift}: x crosses are original data reduced by \citet{1993J_lc}; \textit{Swift} in 2009: filled circles are original in this work.}
\label{SN 1993J X-ray lc Image}
\end{figure}

The early detection suggested that the existence of dense CSM region which was close to progenitor. OSSE on the GRO satellite determined the initial temperature of SN 1993J on the order of 80 keV \citep{Universe_in_X-rays, 1993J_XMM} in the first month. With the extreme high initial temperature and the limitations of the instrument, the ASCA measurements a lower limit of 10 keV was reported \citep{1993J_XMM}. In April 2001, SN 1993J was observed with both the PN camera of the \textit{XMM-Newton} observatory for exposing about 70 ks. The \textit{XMM-Newton} spectrum was fitted by double-\textit{vmekal}, a 2-component thermal model. The best fit indicated that the shock wave temperatures were $0.34^{+0.05}_{-0.03}$ and $6.54^{+4}_{-4}$ keV respectively \citep{1993J_XMM} after about 2900 days. 

\begin{figure}
\includegraphics[width=84mm]{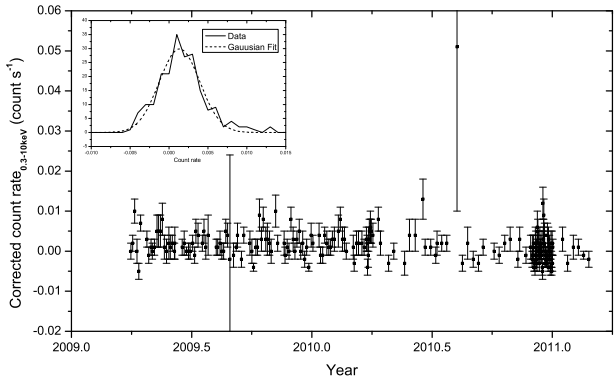}
\caption{\textit{Swift} X-ray lightcurve of energy band 0.3--10 keV of SN 1993J in 2009. The curve is flat and the count rate is low compare with the former years. There was a short bump in the X-ray light curve after day 6025 which shows the uneven distribution of the CSM.}
\label{SN 1993J X-ray lc Image 2009}
\end{figure}

Unlike the unusual X-ray evolution of SN 1987A with a rapid enhancement of X-ray flux. As shown in the Fig. \ref{SN 1993J X-ray lc Image}, the X-ray luminosity of the SN 1993J was decreasing over its life. \citet{1993J_IAU} suggested the luminosity in range of 0.3--2.4 keV decay with $t^{-3}$ over the first 200 days. With the assumptions of the mass loss rate of the progenitor wind can be expressed as
\begin{equation}
\dot{M}=4\pi\rho v_{w}r^{s}.
\end{equation}
and the luminosity is roughly proportional to square of the density, the density profile index was estimated as $s=1.65$ \citep{1993J_XMM}.

The \textit{Swift} data taken between 2009 and 2011 (Fig. \ref{SN 1993J X-ray lc Image 2009}) shows the decay rate of the X-ray flux was slowing down. The light curve is almost a constant ??? or ??? with one sigma fluctuation ??? or ??? in that period. 

\subsection{SN 2007od}
SN 2007od was detected by S. Maticic on four unfiltered CCD images taken around Nov. 2.85 UT with the 60-cm f/3.3 Cichocki reflector in the course of the Comet and Asteroid Search Program (PIKA) at Crni Vrh Observatory. The optical position is at R.A. = 23h55m48s.68, Decl. = +18deg24${\arcmin}$54.8${\arcsec}$ \citep{2007od}. 

SN 2007od was observed by \textit{Swift} initially 3 days after the explosion. No significant X-ray signal was detect at the position in the 4.0 ks XRT exposure by that observation \citep{2007od_no}. After combining with the later observations, there was a X-ray source detected at the SN position with a 5.7$\sigma$ significance in the merged 58.2 ks XRT data. The detected count rate was $(4.8\pm1.2)\times10^{-4}\,\mathrm{count\,s^{-1}}$ which is corresponding to a flux rate of ${1.8\times10^{-14}}$ erg cm${^{-2}}$\,s${^{-1}}$ with a power law model of photon index 2. The reason of the deferred X-ray emission is most likely absorption by the cool, dense shell formed by the reverse shock at the early epoch \citep{inter2, Nymark_2006}. 
\begin{equation}
L_x \propto t^{-(15-6s+sn-2n)/(ns)},
\end{equation}
where $r^{-n}$ is the radial dependence density profile of the SN ejecta and $s$ is the profile index of (4). By taking a reasonable assumption of the progenitor is compact with $n = 10$ \citep{inter_1993J}, it gives $L_x \propto t^{-(4s-5)/(10s)}$. Compare the equation with the observed decline of $t^{-0.6}$, the value $s \approx 2$ which is consistent to the initial interaction of an expanding Type II supernova with a CSM created by mass loss of the progenitor \citep{self-solution}. 


\begin{figure}
\includegraphics[width=84mm]{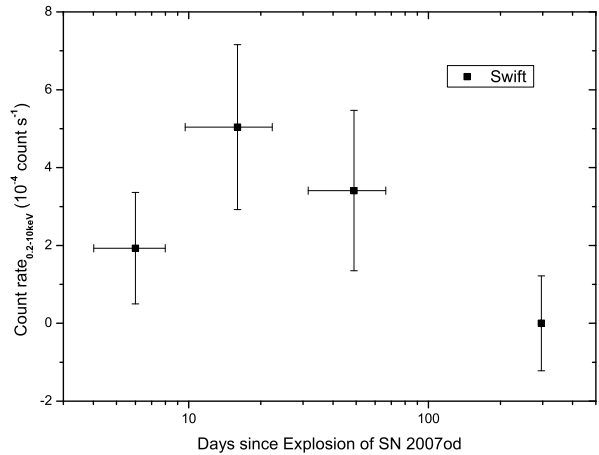}
\caption{\textit{Swift} X-ray lightcurve of SN 2007od in range of 0.3 to 10 keV.}
\label{SN 2007od X-ray lc Image}
\end{figure}

\section{Discussions and Conclusions}
We have investigated over 6 years archival \textit{Swift} observations taken from November 2004 to February  2011 and discovered three new candidates (i.e. SN 1986L, 2003lx and 2007od) out of 24 X-ray detectable \textit{Swift} SNe. It is considerable because there were only $\sim40$ detections of X-ray SNe in the past, we recovered half of them and increased the total detections by $\sim10$\%. Furthermore, The automatic selection program developed in this project worked well to select useful supernova data from over thirty thousand observations in the archive. The success of this work ensures the feasibility of other analogous surveys in \textit{Swift}. It is also promising that more X-ray SNe detections would be found if this complete survey lasts for a few more years. 

There is no doubt that every supernovae has its unique features. Even the supernovae of the same type, the variations between environments, stellar abundances, stellar wind and shock wave properties are doubtless to influence the X-ray emissions. It is difficult to find a simply model to satisfy all features of all X-ray SNe spectra perfectly. However, in a macroscopic view, it is still fair to say that most of the X-ray SNe spectra are similar to each other. Power law model matches most of the cases may be a tangible testimony of that. Based on this thought, we have done several statistics on their luminosities, hardness ratios and the CSM density profile. The results show that most of the SN luminosities lie on the $10^{39}$ to $10^{41}\mathrm{\,erg\,s{^{-1}}}$. The studies of hardness ratios and CSM density profile partly ensure the theoretical models \citep{fransson_82,inter2}.

One of the important contributions of this work is the discovery of the Type Ia X-ray SN candidate, SN 2003lx. The X-ray luminosity was $\sim{4.8\times10^{41}}\mathrm{\,erg\,s{^{-1}}}$ in 0.3--2.0 keV which was inferred by power law of index $\sim 2$ and distance 155 Mpc. The luminosity is high compare to the other Ia SNe which have typical luminosity of a upper limit ${\sim 10^{38-39}}\mathrm{\,erg\,s{^{-1}}}$ \citep{2005ke,chan_typeIsn}. However, observations taken by ROSAT in 1990 provides a strong evidence that there was no X-ray detection at the SN position before the SN explosion which is consistent to a Type Ia SN emission. 

We did some study on the SN 1987A, 1993Jx to monitor the their evolutions. Basic analysis of the three new SNe were done to see their X-ray properties. In fact, there are still several X-ray SNe potential candidates (for example, SN 1989B and SN 2007ax) which did not show in this paper as their X-ray signals are weak and/or the off-set from SN position are huge. More Observations or better images (by \textit{Chandra} or XMM-\textit{Newton}) are expected to confirm these potential candidates.

\section*{Acknowledgments}

We thank Dr. Albert K.H. Kong and his postgraduates from National Tsing Hua University for engaging and helpful discussions. This work is based on observations obtained with \textit{Swift}, a part of NASA's medium explorer program, with the hardware being developed by an international team from the United States, the United Kingdom and Italy, with additional scientific involvement by France, Japan, Germany, Denmark, Spain, and South Africa. K.L. Li acknowledges support from the University of Hong Kong under the giant of Postgraduate Studentship. C.S.J. Pun acknowledges support of a RGC grant from the government of Hong Kong SAR.

\label{lastpage}

\end{document}